\documentclass[a4paper,12pt, epsfig]{article}
\def\letter{0}\def\pr{0}
\pdfoutput=1 
\usepackage{epsfig}
\usepackage{epstopdf}
\usepackage{graphicx}
\usepackage{ifthen}
\usepackage{ulem}

\usepackage{amsmath}
\usepackage[psamsfonts]{amssymb}
\usepackage{euscript}

\usepackage{latexsym}
\usepackage[arrow,matrix,curve]{xy}

\usepackage[hypertexnames=false]{hyperref}
\hypersetup{
    colorlinks=true,
    linkcolor=blue,
    filecolor=magenta,   
    citecolor=black,   
    urlcolor=cyan,
    }

\newskip\humongous \humongous=0pt plus 1000pt minus 1000pt

\newif\ifdtup

\def\,{\hspace{-.1cm}}
\def\hsp{,\hspace{.7cm}}

\def\kis {k_I^{\rm{S}}}
\def\kia {k_I^{\rm{aS}}}
\def\oks {\omega_{{\kis}}}
\def\oka {\omega_{{\kia}}}
\def\sl {{\sqrt{\lambda}}}

\def\fc#1#2 {\frac{n}{q}#1\frac{n}{q}#2}

\def\lim#1{\stackrel{\rm{lim}}{{}_{#1}}}

\newcommand{\vac}{\ensuremath{|0\rangle}}

\renewcommand{\sin}{\textrm{sin}}

\renewcommand{\sinh}{\textrm{sinh}}

\renewcommand{\tanh}{\textrm{tanh}}
\newcommand{\sech}{\textrm{sech}}
\newcommand{\csch}{\textrm{csch}}

\renewcommand{\theequation}{\arabic{section}.\arabic{equation}}
\renewcommand{\(}{\begin{equation}}
\renewcommand{\)}{end{equation} \vspace{-.05in}\linebreak}

\newcounter{saveeqn}
\newcounter{savealpheqn}

\newcommand{\alpheqn}{\setcounter{saveeqn}{\value{equation}}%
  \stepcounter{saveeqn}\setcounter{equation}{0}%
  \renewcommand{\theequation}{\mbox{\arabic{section}.\arabic{saveeqn}
\alph{equation}}}
  \renewcommand{\)}{\end{equation}}}
\def\part#1{\frac{\partial}{\partial{#1}}}%
\def\group#1{\refstepcounter{equation}\setcounter{saveeqn}
 {\value{equation}}%
  \label{#1}\setcounter{equation}{0}%
\renewcommand{\theequation}{\mbox{\arabic{section}.\arabic{saveeqn}
\alph{equation}}}
  \renewcommand{\)}{\end{equation}}}
\newcommand{\reseteqn}{\setcounter{equation}{\value{saveeqn}}%
  \renewcommand{\theequation}{\arabic{section}.\arabic{equation}}%
  \renewcommand{\)}{\end{equation}}}

\newcommand{\aalpheqn}{\setcounter{saveeqn}{\value{equation}}%
  \stepcounter{saveeqn}\setcounter{equation}{0}%
  \renewcommand{\theequation}{\mbox{
        \Alph{subsection}.\arabic{saveeqn}\alph{equation}}}
   \renewcommand{\)}{\end{equation}}}
\newcommand{\areseteqn}{\setcounter{equation}{\value{saveeqn}}%
  \renewcommand{\theequation}{\Alph{subsection}.\arabic{equation}}%
  \renewcommand{\)}{\end{equation}}}

\renewcommand{\thefootnote}{\alph{footnote}}
\renewcommand{\(}{\begin{equation}}
\renewcommand{\)}{\end{equation}}
\newcommand{\ba}{\begin{eqnarray}}
\newcommand{\ea}{\end{eqnarray}}
\renewcommand{\a}{\alpha}
\renewcommand{\b}{\beta}

\newcommand{\cbp}{\mathop{\vtop{\ialign{##\crcr
   $\hfil\displaystyle{}\hfil$\crcr\noalign{\kern-13pt\nointerlineskip}
   \BIG{)}\hskip0pt\crcr\noalign{\kern3pt}}}}}
\newcommand{\pa}{\mathop{\vtop{\ialign{##\crcr

$\hfil\displaystyle{\oplus}\hfil$\crcr\noalign{\kern+1pt\nointerlineskip
}
   \hspace{.08in}$^{\alpha=0}$\hskip6pt\crcr\noalign{\kern3pt}}}}}
\renewcommand{\hsp}{,\hspace{.3in}}
\newcommand{\p}{^\prime}

\catcode`\@=11
\def\vereq#1#2{\lower3pt\vbox{\baselineskip1.5pt \lineskip1.5pt
\ialign{$\m@th#1\hfill##\hfil$\crcr#2\crcr\sim\crcr}}}
\catcode`\@=12

\renewcommand{\(}{\begin{equation}}
\renewcommand{\)}{\end{equation}}

\def\pin#1{\int \frac{d#1}{2\pi}}
\def\ppin#1{\int\hspace{-17pt}\sum \frac{d#1}{2\pi}}

\def\dint{\int\hspace{-12pt}\sum }
\def\pink#1{\int \frac{d^{#1}k}{(2\pi)^{#1}}}

\def\Bd#1{B^\ddag_{k_{#1}}}

\def\cc{\mathcal{C}}
\def\df{\mathcal{D}_{f}}

\def\os{\omega_S}

\def\blu#1{\textcolor{blue}{Jarah: #1}}
\def\red#1{\textcolor{red}{Hui: #1}}

\newcommand{\beas}{\begin{eqnarray*}}
\newcommand{\eeas}{\end{eqnarray*}}

\newcommand{\bquo}{\begin{quote}}
\newcommand{\enqu}{\end{quote}}

\newcommand{\R}{{\mathbb R}}

\newcommand{\g}{{\mathfrak g}}

\def\ok#1{\omega_{k_{#1}}}

\def\V#1{V^{(#1)}(\sqrt{\lambda}f(x))}

\def\ck{\csch\left(\frac{\pi k}{2\b}\right)}

\def\mb{\mathcal{B}}
\def\mc{\mathcal{C}}
\def\md{\mathcal{D}}
\def\me{\mathcal{E}}

\newcommand{\beq}{\begin{equation}}
\newcommand{\eeq}{\end{equation}}
\newcommand{\bea}{\begin{eqnarray}}
\newcommand{\eea}{\end{eqnarray}}

\newskip\humongous \humongous=0pt plus 1000pt minus 1000pt

\newif\ifdtup

\catcode`\@=11

\ifthenelse{\equal{\letter}{0}}{ 

\@addtoreset{equation}{section}
\def\theequation{\arabic{section}.\arabic{equation}}

\def\@normalsize{\@setsize\normalsize{15pt}\xiipt\@xiipt
\abovedisplayskip 14pt plus3pt minus3pt%
\belowdisplayskip \abovedisplayskip
\abovedisplayshortskip \z@ plus3pt%
\belowdisplayshortskip 7pt plus3.5pt minus0pt}

\def\small{\@setsize\small{13.6pt}\xipt\@xipt
\abovedisplayskip 13pt plus3pt minus3pt%
\belowdisplayskip \abovedisplayskip
\abovedisplayshortskip \z@ plus3pt%
\belowdisplayshortskip 7pt plus3.5pt minus0pt
\def\@listi{\parsep 4.5pt plus 2pt minus 1pt
      \itemsep \parsep
      \topsep 9pt plus 3pt minus 3pt}}

\relax

\def\section{\@startsection{section}{1}{\z@}{3.5ex plus 1ex minus  .2ex}{2.3ex plus .2ex}{\large\bf}}

\def\thesection{\arabic{section}}
\def\thesubsection{\arabic{section}.\arabic{subsection}}

\def\appendix{\setcounter{section}{0}
 \def\thesection{Appendix \Alph{section}}
 \def\thesubsection{\Alph{section}.\arabic{subsection}}
 \def\theequation{\Alph{section}.\arabic{equation}}}
\renewcommand{\theequation}{\arabic{section}.\arabic{equation}}

}{
\renewcommand{\theequation}{\arabic{equation}}

} 

\begin{document}
\def\thefootnote{\fnsymbol{footnote}}
\def\thetitle{(Anti-)Stokes Scattering on Kinks}
\def\auttwo{Hui Liu}
\def\autone{Jarah Evslin}
\def\affc{School of Fundamental Physics and Mathematical Sciences, Hangzhou Institute for Advanced Study,
University of Chinese Academy of Sciences, Hangzhou 310024, China}
\def\affb{University of the Chinese Academy of Sciences, YuQuanLu 19A, Beijing 100049, China}
\def\affd{Arnold Sommerfeld Center, Ludwig-Maximilians-Universität, Theresienstraße 37, 80333 München, Germany}
\def\affa{Institute of Modern Physics, NanChangLu 509, Lanzhou 730000, China}
\def\affe{Institute of Contemporary Mathematics, School of Mathematics and Statistics,
Henan University, Kaifeng, Henan 475004, P. R. China}


\ifthenelse{\equal{\pr}{1}}{
\title{\thetitle}
\author{\autone}
\author{\auttwo}
\author{\autthree}
\affiliation {\affa}
\affiliation {\affb}

}{

\begin{center}
{\large {\bf \thetitle}}

\bigskip

\bigskip


{\large \noindent  \autone{${}^{1,2}$} \footnote{jarah@impcas.ac.cn} 
and \auttwo{${}^{3,2,4}$} \footnote{hui.liu@campus.lmu.de}
}


\vskip.7cm

1) \affa\\
2) \affb\\
3) \affc\\
4) \affd\\

\end{center}

}

\begin{abstract}
\noindent
At leading order, there are three inelastic scattering processes beginning with a quantum kink and a fundamental meson.  Meson multiplication, in which the final state is a kink and two mesons, was treated recently.  In this note we treat the other two, (anti)-Stokes scattering, in which the kink's shape mode is (de-)excited and the final state contains one meson.  In the case of a general scalar kink, we find analytic formulas for the forward and backward scattering amplitudes and probabilities as functions of the momentum of the incident meson.  The general results are then specialized to  the kink of the $\phi^4$ double-well model.


\end{abstract}

%
\setcounter{footnote}{0}
\renewcommand{\thefootnote}{\arabic{footnote}}

\ifthenelse{\equal{\pr}{1}}
{
\maketitle
}{}

\section{Introduction}

Scalar theories in (1+1)-dimensions provide some of the simplest quantum field theory models.  If the scalar field is subjected to a degenerate potential, in addition to fundamental meson excitations there will also be nonperturbative kinks.  

The classical field theory of such models is already surprisingly rich.  Most attention has focused on kink-antikink scattering \cite{kklan,kk0,kk1,kk2}.  Here, it has long been known \cite{csw} that the range of initial relative speeds leading to distinct outcomes has a fractal structure of resonance windows.  It was once thought that this results from the internal excitation spectrum of the kinks.  Certainly they play a role \cite{chaosmod,int,int2}.  However it was then found \cite{doreyf6,f622} that such windows appear even in models in which the kink has no internal excitations.  Thus, it has become clear that the interactions of kinks with the bulk dynamics are important \cite{sfal21,col22}.

This bulk dynamics is itself quite rich.  There are spectral walls \cite{muri,fmuri} beyond which internal excitations of kinks escape into the continuum.  While these have striking consequences classically, in the quantum theory they are rather smooth \cite{chris}.  Also, after a kink-antikink collision, many kinds of modes are excited.  In the end, only the longest-lived remain.  Among these, the oscillons \cite{osc,osc3d} survive for an amazingly long time.  However, again the quantum theory appears to be different.  In the quantum theory, new decay channels open which greatly reduce the oscillon lifetime \cite{quantosc}.

In summary, two critical pieces of the picture are missing. The first is a systematic understanding of interactions between kinks and bulk degrees of freedom.  The second is an understanding of the role played by quantum corrections, and whether they disappear in the classical limit or, as the oscillon case and perhaps the spectral wall case seem to suggest, radically affect the physics.

This motivates an understanding of interactions between kinks and elementary meson quanta in the quantum theory.  Such interactions are much simpler than kink-antikink interactions.  However, the classic literature on quantum kink-meson scattering has been limited largely to finding effective Yukawa couplings between kinks and mesons \cite{hayashi1,hayashi2}.  

Recently, a linearized perturbation theory for such models has been developed at one-loop in Ref.~\cite{mekink} and beyond in Ref.~\cite{me2loop}.  It greatly simplifies calculations in the one-kink sector, which consists of states with a kink and any finite number of mesons, with respect to the traditional collective coordinate approach of Refs.~\cite{gjscc,gj76}.  

Using this approach it was soon realized that, at leading order, there are precisely three inelastic scattering processes which begin with a single kink and a single meson.  The first is meson multiplication, in which the meson is absorbed by the kink and two mesons are emitted.  This was recently studied in Ref.~\cite{memult}.  The other two are Stokes and anti-Stokes scattering.  Stokes scattering is a process where a meson scatters off of a ground state kink while exciting its shape mode.  Anti-Stokes scattering is a process where a meson scatters off of an excited kink and deexcites its shape mode.  

In the present note, we present the first-ever treatment of these two processes.  After a review of linearized kink perturbation theory in Sec.~\ref{revsez}, we calculate the probability, as a function of the momentum of the incoming meson, of Stokes and anti-Stokes scattering in Secs.~\ref{stokessez} and \ref{asez} respectively.  Our results are specialized to the $\phi^4$ double-well model in Sec.~\ref{fsez}.

\section{Linearized Kink Perturbation Theory} \label{revsez}

\subsection{The Main Idea}

In Refs.~\cite{mekink,me2loop} a new, Hamiltonian formalism has been introduced for calculations in the kink sector of a quantum theory of a scalar field $\phi(x)$ and its conjugate $\pi(x)$ in 1+1 dimensions.  The kink sector is the Fock space of states consisting of a finite number of fundamental mesons in addition to a single quantum kink.  We will refer to the Fock space of mesons in the absence of a kink as the vacuum sector.

The vacuum sector states can be constructed in perturbation theory.  One decomposes the field in a plane wave basis, constructing creation and annihilation operators.   The vacuum is defined as the state which is annihilated by all annihilation operators, and the vacuum sector is generated by finite numbers of creation operators acting on the vacuum.  Hamiltonian eigenstates can be found perturbatively by solving the Hamiltonian eigenvalue problem.

This perturbative approach fails for the kink sector.  This is evident already in classical theory, where it results from the fact that large moments of the field do not tend to zero.  The kink sector corresponds to classical field configurations which are close to the classical kink solution $\phi(x,t)=f(x)$.  The higher moments of $\phi(x,t)-f(x)$ are small, and so one expects a perturbative approach in $\phi(x,t)-f(x)$ to yield kink sector states.

Linearized kink perturbation theory is a formalism for doing this in quantum field theory.  A unitary displacement operator $\df$ is constructed in the Schrodinger picture as
\beq
\df={{\rm Exp}}\left[-i\int dx f(x)\pi(x)\right].
\eeq
We use $\df$ as a passive transformation, renaming the coordinate system of the Hilbert space and transforming the operators that act on them.  More precisely, we define the {\it{kink frame}} as the coordinate system on the Hilbert space in which the ket $|\psi\rangle$ represents the state $\df|\psi\rangle$ as defined in the usual, defining frame.  With this definition, it is easily shown that, in the kink frame, energies are measured and time is evolved by the kink Hamiltonian $H\p$
\beq
H\p=\df^\dag H\df \label{df}
\eeq
where $H$ is the original Hamiltonian, which defines the theory.  

What have we gained?  In the kink frame, it is the kink sector which is constructed perturbatively using creation operators.  Thus in the presence of a kink, the construction of Hamiltonian eigenstates, form factors, and even amplitudes and probabilities for various processes are reduced to perturbative problems in the kink frame.

What have we lost?  We had to choose a particular kink solution $f(x)$.  In a translation-invariant theory, there would be a moduli space of choices $f(x-x_0)$ for every real $x_0$.  Thus we have lost manifest translation invariance.  We must work locally, close to some base point in moduli space.  However, if we are interested in translation-invariant states, or more precisely momentum eigenstates, which we will be in this paper\footnote{This paper will be entirely in the center of mass frame of the kink and meson, and so all states will be eigenstates of the total momentum operator with eigenvalue zero.  Wave packets will be constructed consisting of different momenta for the meson, recalling that the kink momenta will always be equal and opposite.  Translation-invariance is with respect to simultaneous and equal translations of the kink and mesons.}, then it is sufficient to understand any region in moduli space to understand every region.  And so this will not be a problem, we simply work perturbatively in $x_0$, and impose  translation-invariance at will to simplify expressions.  

\subsection{The Details}

While we expect this formalism to apply quite generally, so far we have only applied it to Schrodinger picture Hamiltonians of the form
\begin{equation}
H=\int d x: \mathcal{H}(x):_a, \quad \mathcal{H}(x)=\frac{\pi^2(x)}{2}+\frac{\left(\partial_x \phi(x)\right)^2}{2}+\frac{V(\sqrt{\lambda} \phi(x))}{\lambda}.
\end{equation}
Here $V$ is a degenerate potential and $\phi(x,t)=f(x)$ is a solution of the classical equations of motion which interpolates between two degenerate minima.

The notation $::_a$ represents normal ordering of the creation and annihilation operators for plane waves.  It is defined at a mass scale $m$, which has two definitions
\beq
m^2=V^{(2)}(\sqrt{\lambda} f(\pm \infty))\hsp
V^{(n)}(\sqrt{\lambda} \phi(x))=\frac{\partial^n V(\sqrt{\lambda} \phi(x))}{(\partial \sqrt{\lambda} \phi(x))^n}
\eeq
corresponding to the scalar mass at the two minima of the potential at infinity.  If these disagree, then quantum corrections break the degeneracy and the kink becomes an accelerating false vacuum bubble wall \cite{wstabile}.  We will not consider this case.

We find the eigenstates of the kink Hamiltonian $H\p$ perturbatively in the coupling constant $\lambda$.  To do this we decompose all quantities in powers of $\lambda$.  For example, the energy $Q$ of the kink ground state is decomposed into $\sum_i Q_i$ where each $Q_i$ is of order $O(\lambda^{i-1})$.  Note that $Q_0$ is just the classical energy of the classical kink solution.

The kink Hamiltonian itself is decomposed into terms $H\p_i$ with $i$ factors of the fundamental fields, when normal ordered.  These include
\beq
H\p_0=Q_0\hsp H\p_1=0\hsp
H\p_{n>2}=\lambda^{\frac{n}{2}-1}\int dx \frac{V^{(n)}(\sqrt{\lambda} f(x))}{n !}: \phi^n(x):_a.\label{hn}
\eeq

The most important is $H\p_2$, as its eigenvectors are the first step in the perturbative expansion for the kink sector states.  Written in terms of $x$ it is rather odd, resembling a free Hamiltonian but with a position-dependent mass term.  

To write it more transparently, we will introduce the kink's normal modes $\g(x)$, defined to be small, classical fluctuations about the kink, which solve the Sturm-Liouville equation
\beq
\V{2}{\g}(x)=\omega^2{\g}(x)+{\g}^{\prime\prime}(x)\hsp \phi(x,t)=e^{-i\omega t}\g(x). \label{sl}
\eeq
They are classified by their frequency $\omega$.  The real solution $\g_B(x)$ with $\omega_B=0$ is called the zero mode.  Any real mode $\g_S(x)$ with $0<\os<m$ is called a shape mode.  Above this lie the continuum modes $\g_k(x)$ with $\ok{}=\sqrt{m^2+k^2}$.  We fix the conventions
\bea
\ok{}&=&\sqrt{m^2+k^2}\hsp \g^*_k(x)=\g_{-k}(x)\label{comp}\\
\int dx |{\g}_{B}(x)|^2&=&1,\
\int dx {\g}_{k_1} (x) {\g}^*_{k_2}(x)=2\pi \delta(k_1-k_2),\ 
\int dx {\g}_{S_1}(x){\g}^*_{S_2}(x)=\delta_{S_1S_2}. \nonumber
\eea

The normal modes generate all bounded functions, and so, instead of plane waves, we may use them to decompose the fields \cite{cahill76}
\bea
\phi(x) &=&\phi_0 \mathfrak{g}_B(x)+\ppin{k} \left(B_k^{\ddag}+\frac{B_{-k}}{2 \omega_k}\right) \mathfrak{g}_k(x) \hsp
B_k^{\ddagger}=\frac{B_k^{\dagger}}{\left(2 \omega_k\right)}\hsp
B_S^{\ddagger}=\frac{B_S^{\dagger}}{\left(2 \omega_S\right)}
\label{dec}\\
\pi(x) &=&\pi_0 \mathfrak{g}_B(x)+i \ppin{k}\left(\omega_k B_k^{\ddag}-\frac{B_{-k}}{2}\right) \mathfrak{g}_k(x) \hsp
B_{-S}=B_S\hsp
\ppin{k}=\pin{k}+\sum_S
\nonumber
\eea
into operators $\phi_0,\ \pi_0,\ B$\ and $B^\ddag$.  This provides a new basis of our operator algebra, and any operator may be written in terms of these operators.  The canonical commutation relations satisfied by $\phi(x)$ and $\pi(x)$ imply that these satisfy
\beq
\left[\phi_0, \pi_0\right]=i, \quad\left[B_{S_1}, B_{S_2}^{\ddagger}\right]=\delta_{S_1 S_2}, \quad\left[B_{k_1}, B_{k_2}^{\ddagger}\right]=2 \pi \delta\left(k_1-k_2\right).
\eeq

Finally we are ready to write $H_2\p$.  It is \cite{cahill76}
\begin{equation}
H\p_2=Q_1+H_{\text {free }}, \quad H_{\text {free }}=\frac{\pi_0^2}{2}+\sum_S\omega_S B_S^{\ddag} B_S+\int \frac{d k}{2 \pi} \omega_k B_k^{\ddag} B_k. \label{h2}
\end{equation}
Here $Q_1$ is the one-loop correction to the kink mass.  The $\pi_0^2$ term is the kinetic energy of a free quantum-mechanical particle  of mass $Q_0$ with position operator $\phi_0/\sqrt{Q_0}$.  This particle is the center of mass of the kink.  The other terms are quantum harmonic oscillators for the shape modes $S$ and continuum modes $k$.  $B^\ddag_S$ excites a shape mode, while $\Bd{}$ excites a continuum mode.

The kink ground state $\vac$ of the kink Hamiltonian $H\p$ can be decomposed into contributions $\vac_i$, of order $O(\lambda^{i/2}).$  The first term in our semiclassical expansion, $\vac_0$, is the vacuum of $H\p_2$.  It is the ground state of each term in (\ref{h2}), and so is completely characterized by the conditions
\beq
\pi_0\vac_0=B_k\vac_0=B_S\vac_0=0. \label{v0}
\eeq

\section{Stokes Scattering} \label{stokessez}

In a Stokes scattering event, one meson is absorbed by a ground state kink, one meson is emitted and a shape mode is excited.  The initial condition is therefore a superposition 
\bea
|\Phi\rangle_0&=&\pin{k_1}\a_{k_1}|k_1\rangle_0\hsp
\alpha_k=\int d x \Phi(x) \mathfrak{g}_k(x)\label{pach}\\
\Phi(x)&=&\operatorname{Exp}\left[-\frac{\left(x-x_0\right)^2}{4 \sigma^2}+i x k_0\right], \quad x_0 \ll-\frac{1}{ m}, \quad  \frac{1}{k_0},\frac{1}{m}\ll\sigma \ll\left|x_0\right|\nonumber
\eea
of one-meson states
\beq
|k_1\rangle_0=\Bd 1\vac_0
\eeq
in the kink sector.  Here the meson wave packet begins at $x=x_0$, which is far to the left of the kink, which is at $x=0$.  It moves to the right with momentum roughly equal to $k_0$.  The final state consists of a meson and a kink whose shape mode is excited.  It is therefore a superposition of states of the form
\beq
|Sk_2\rangle_0=B^\ddag_S\Bd 2\vac_0. \label{fs}
\eeq

At lowest order, $O(\sqrt{\lambda})$, the only term in the kink Hamiltonian that can interpolate between these states is
\bea
H_I&=&\frac{\sqrt{\lambda}}{2} \int \frac{d k_1}{2 \pi} \frac{d k_2}{2 \pi}  \frac{V_{S,k_2,-k_1}}{\omega_{k_1}}  B_{S}^{\ddagger}B_{k_2}^{\ddagger} B_{k_1} \\
V_{S,k_2,-k_1}&=&\int d x V^{(3)}(\sqrt{\lambda} f(x)) \mathfrak{g}_{S}(x) \mathfrak{g}_{k_2}(x) \mathfrak{g}_{-k_1}(x) .\nonumber
\eea






At order $O(\sl)$, the corresponding terms in the time evolution operator are
\beq
e^{-it(H\p_2+H_I)}=e^{-itH\p_2}-i\int _0^t dt_1 e^{-i(t-t_1)H\p_2}H_I e^{-i t_1 H\p_2}+O(\lambda).
\eeq
We will drop the first term, as it will not contribute to the matrix elements below.  Acting this on a one-kink, one-meson state one finds Stokes scattering
\beq
e^{-iH\p t}|k_1\rangle_0\bigg|_{O(\sl)}=\frac{-i\sqrt{\lambda}}{2\omega_{k_1}} \pin{k_2}V_{S,k_2,-k_1}e^{-\frac{it}{2}(\omega_{k_1}+\os+\ok{2})}\frac{{\rm sin}\left[\left(\frac{\os+\ok{2}-\ok{1}}{2}\right)t\right]}{(\os+\ok{2}-\ok{1})/2}|S k_2\rangle_0.
\eeq

This process is on-shell when $k_1=\pm {\kis}$ where we have defined
\beq
{\oks}=\ok{2}+\os\hsp {\kis}>0.
\eeq
At large times, we may use the identity
\beq
\lim{t\rightarrow\infty}\frac{{\rm sin}\left[\left(\frac{\os+\ok{2}-\ok{1}}{2}\right)t\right]}{(\os+\ok{2}-\ok{1})/2}=2\pi\delta(\os+\ok{2}-\ok{1})=\left(\frac{{\oks}}{{\kis}}\right)\left(2\pi\delta(k_1-{\kis})+2\pi\delta(k_1+{\kis})\right)
\eeq
to perform the $k_2$ integral.  Folding this result into the wave packet (\ref{pach}), one finds the Stokes scattered part of the state at large times $t$
\bea
e^{-iH\p t}|\Phi\rangle_0\bigg|_{O(\sl)}&=&-i\sqrt{\lambda}\pin{k_1}\frac{\alpha_{k_1}}{2\ok{1}} \pin{k_2}V_{S,k_2,-k_1}e^{-\frac{it}{2}(\ok{1}+\os+\ok{2})}\frac{{\rm sin}\left[\left(\frac{\os+\ok{2}-\ok{1}}{2}\right)t\right]}{(\os+\ok{2}-\ok{1})/2}|S k_2\rangle_0\nonumber\\
&=&\frac{-i\sqrt{\lambda}}{2} \pin{k_2}\frac{e^{-i{\oks} t}}{{\kis}}\left(\alpha_{{\kis}}V_{S,k_2,-{\kis}}+\alpha_{-{\kis}}V_{S,k_2,{\kis}}
\right)|S k_2\rangle_0. \label{sev}
\eea

The meson wave packet begins far from the kink, where one may apply the asymptotic form of the normal modes
\bea
\g_k(x)&=&\left\{\begin{tabular}{lll}
$\mb_ke^{-ikx}+\mc_ke^{ikx}$&\rm{if} & $x\ll  -1/m$\\
$\md_ke^{-ikx}+\me_k e^{ikx}$&\rm{if} & $x\gg 1/m$\\
\end{tabular}
\right. \label{gk}\\
|\mb_k|^2+|\mc_k|^2&=&|\md_k|^2+|\me_k|^2=1\hsp
\mb^*_k=\mb_{-k}\hsp
\mc^*_k=\mc_{-k}\hsp
\md^*_k=\md_{-k}\hsp
\me^*_k=\me_{-k}\nonumber
\eea
to evaluate the coefficients $\alpha_k$ of the wave packet.  As $\kis$ is defined to be positive and $k_0$ is chosen to be positive, in Eq.~(\ref{sev}) only two cases appear
\bea
\alpha_{{\kis}}&=&2\sigma\sqrt{\pi}\left[\mb_{{\kis}}e^{ix_0(k_0-{\kis})}e^{-\sigma^2(k_0-{\kis})^2}+\mc_{{\kis}}e^{ix_0(k_0+{\kis})}e^{-\sigma^2(k_0+{\kis})^2}
\right]\\
&=&2\sigma\sqrt{\pi}\mb_{{\kis}}e^{ix_0(k_0-{\kis})}e^{-\sigma^2(k_0-{\kis})^2}\nonumber
\eea
and
\bea
\alpha_{-{\kis}}&=&2\sigma\sqrt{\pi}\left[\mb^*_{{\kis}}e^{ix_0(k_0+{\kis})}e^{-\sigma^2(k_0+{\kis})^2}+\mc^*_{{\kis}}e^{ix_0(k_0-{\kis})}e^{-\sigma^2(k_0-{\kis})^2}
\right]\\
&=&2\sigma\sqrt{\pi}\mc^*_{{\kis}}e^{ix_0(k_0-{\kis})}e^{-\sigma^2(k_0-{\kis})^2}.\nonumber
\eea
Substituting these back into Eq.~(\ref{sev}), one finds the relevant part of the state at large times $t$
\bea
e^{-iH\p t}|\Phi\rangle_0\bigg|_{O(\sl)}&=&-i\sigma\sqrt{\pi\lambda} \pin{k_2}e^{ix_0(k_0-{\kis})}e^{-\sigma^2(k_0-{\kis})^2}e^{-i{\oks}t}\left(\frac{\tilde{V}_{S,k_2,-{\kis}}}{{\kis}}\right)|S k_2\rangle_0\nonumber\\
\tilde{V}_{S,k_2,-{\kis}}&=&\mb_{{\kis}}V_{S,k_2,-{\kis}}+\mc^*_{{\kis}}V_{S,k_2,{\kis}}.
\eea
Note that in the case of a reflectionless kink, $\mc=0$ and so $\left|\tilde{V}\right|=\left|V\right|$.

Using the inner product
\beq
{}_0\langle S k_1|S k_2\rangle_0=\frac{2\pi\delta(k_1-k_2)}{4\os\ok{2}}{}_0\langle 0\vac_0
\eeq
we find the matrix elements
\beq
\frac{{}_0\langle S k_2|e^{-iH\p t}|\Phi\rangle_0}{{}_0\langle 0\vac_0}=\frac{-i\sigma\sqrt{\pi\lambda}}{4\os\ok{2}{\kis}} e^{ix_0(k_0-{\kis})}e^{-\sigma^2(k_0-{\kis})^2}e^{-i{\oks}t}\tilde{V}_{S,k_2,-{\kis}}
\eeq
which square to
\bea
\left|\frac{{}_0\langle S k_2|e^{-iH\p t}|\Phi\rangle_0}{{}_0\langle 0\vac_0}\right|^2&=&\frac{\sigma^2\pi\lambda}{16\os^2\ok{2}^2{\kis}^2}\left|\tilde{V}_{S,k_2,-{\kis}}\right|^2e^{-2\sigma^2(k_0-{\kis})^2}\\
&=&\frac{\sigma\pi^{3/2}\lambda}{16\sqrt{2}\os^2\ok{2}^2{\kis}^2}\left|\tilde{V}_{S,k_2,-{\kis}}\right|^2\delta({\kis}-k_0).\nonumber
\eea
The last equality holds in the limit $\sigma\rightarrow\infty$.

To calculate the Stokes scattering probability, we will need the projector $\mathcal{P}$ onto final states with an excited kink and a single meson
\beq
\mathcal{P}=\int dk_2 \mathcal{P}_{\rm{diff}}(k_2)\hsp
\mathcal{P}_{\rm{diff}}(k_2)=\frac{4\os\ok{2}}{2\pi} \frac{|Sk_2\rangle_0 {}_0\langle Sk_2|}{{}_0\langle 0\vac_0} .
\eeq


Using the inner product
\beq
\frac{{}_0\langle k_1|k_2\rangle_0}{{}_0\langle 0\vac_0}=\frac{2\pi\delta(k_1-k_2)}{2\ok{1}}
\eeq
one obtains the normalization of the initial state
\bea
\frac{{}_0\langle\Phi|\Phi\rangle_0}{{}_0\langle 0\vac_0}&=&\pink{2}\alpha_{k_1}\alpha^*_{k_2}\frac{{}_0\langle k_2|k_1\rangle_0}{{}_0\langle 0\vac_0}=\pin{k}\frac{|\alpha_k|^2}{2\ok{}}=\frac{1}{2\omega_{k_0}}\pin{k}|\alpha_k|^2\\
&=&\frac{1}{2\omega_{k_0}}\pin{k}\int dx\int dy g_k(x)g^*_k(y)\Phi(x)\Phi^*(y)
\nonumber\\
&=&\frac{1}{2\omega_{k_0}}\int dx |\Phi(x)|^2=\frac{\sigma\sqrt{\pi}}{\sqrt{2}\omega_{k_0}}\nonumber
\eea
where we used $\ok{}\sim \ok{0}$ in the last step in the first line.

Both ${}_0\langle k_1|k_2\rangle_0$ and ${}_0\langle 0\vac_0$ are infinite, and so the previous expression is strictly speaking not defined.  In Ref.~\cite{meip} we describe how such inner products may be calculated systematically, by dividing the numerator and denominator by the translation group.  There are corrections with respect to the naive manipulations above, as a result of the nondiagonal action of the translation operator in the kink frame.  However, these corrections are always subleading by a power of $\sl$ and so do not affect our probability at $O(\lambda)$.

Finally we may assemble all of these ingredients to write the total probability of Stokes scattering at $O(\lambda)$
\bea
P_{\rm{S}}&=&\frac{{}_0\langle\Phi|e^{iH\p t}\mathcal{P}e^{-iH\p t}|\Phi\rangle_0}{{}_0\langle\Phi|\Phi\rangle_0}=\pin{k_2} \frac{4\os\ok{2}}{{}_0\langle 0\vac_0}\frac{\left|{}_0\langle Sk_2|e^{-iH\p t}|\Phi\rangle_0\right|^2}{{}_0\langle\Phi|\Phi\rangle_0/{}_0\langle 0\vac_0}\frac{1}{{}_0\langle 0\vac_0}\\
&=&\pin{k_2}4\os\ok{2}\frac{\frac{\sigma\pi^{3/2}\lambda}{16\sqrt{2}\os^2\ok{2}^2{\kis}^2}\left|\tilde{V}_{S,k_2,-{\kis}}\right|^2\delta({\kis}-k_0)}{\left(\frac{\sigma\sqrt{\pi}}{\sqrt{2}\omega_{k_0}}\right)}
\nonumber\\
&=&\frac{\pi\lambda\ok{0}}{4\os (\ok{0}-\os)k_0^2}\pin{k_2}\left|\tilde{V}_{S,k_2,-{\kis}}\right|^2\delta({\kis}-k_0)\nonumber\\
&=&\lambda\frac{\left|\tilde{V}_{S,\sqrt{(\ok{0}-\os)^2-m^2},-k_0}\right|^2+\left|\tilde{V}_{S,-\sqrt{(\ok{0}-\os)^2-m^2},-k_0}\right|^2
}{8\os k_0\sqrt{(\ok{0}-\os)^2-m^2}}.\nonumber
\eea
We see that the probability is the sum of two terms.  The first is the probability that the emitted meson travels in the same direction as the initial meson, while the second is the probability that it travels in the opposite direction.  We will see in an example below that such reflection occurs even in the case of a reflectionless kink. 

In the initial and final states (\ref{pach}) and (\ref{fs}), the meson travels at a constant velocity $k_0/\ok 0$ when far from the kink.  However an order $O(\sl)$ quantum correction to these states, when evolved with respect to $H\p_2$, can in principle contribute to the amplitude at the same order $O(\sl)$ as the leading term in the states when evolved with $e^{-it(H\p_2+H_I)}$ at $O(\sl)$.  Even though our initial and final states (\ref{pach}) and (\ref{fs}) contain no such $O(\sl)$ correction, such a correction would be created by the evolution $e^{-itH\p}$ as the meson travels far from a kink \cite{memult}.  

As described in Ref.~\cite{memult}, one can include an order $O(\sl)$ quantum correction to the initial and final states so that they are undeformed as they travel, while far from the kink.  Such states are eigenstates not of the kink Hamiltonian $H\p$, but rather of the left and right vacuum Hamiltonians, which are defined by expanding the defining Hamiltonian about the vacua to the left and right of the kink.  These quantum corrections arise from the three-meson vertex far from the kink.  Far from the kink, the mesons separately conserve momentum.  As a result, these processes are far off-shell, leading to a cloud of far off-shell mesons about the initial and final mesons.  One therefore expects that this cloud does not contribute to the asymptotic probability of meson multiplication or Stokes scattering.  In Ref.~\cite{memult} it was shown, in the case of meson multiplication, that this is indeed the case.  The argument proceeds identically here, as Stokes scattering is just meson multiplication in which one of the created mesons is a bound state.  Therefore we conclude that there are also no initial or final state corrections here.

\section{Anti-Stokes Scattering} \label{asez}

In anti-Stokes scattering, the kink begins with an excited shape mode and an approaching meson wave packet.  The initial state is thus
\beq
\left|\Phi\right\rangle_0=\pin{k_1} \alpha_{k_1}\left|S k_1\right\rangle_0\hsp
|S k_1\rangle_0=B^\ddag_S \Bd 1 \vac_0. \label{inita}
\eeq
The final state consists of a meson packet and a deexcited kink,  and so is in the space of states spanned by $|k_2\rangle_0$.

At $O(\sl)$, the only term which interpolates between these two states is
\beq
H_I=\frac{\sqrt{\lambda}}{4\os} \int \frac{d k_1}{2 \pi} \frac{d k_2}{2 \pi}  \frac{V_{S,k_2,-k_1}}{\omega_{k_1}} B_{k_2}^{\ddagger} B_{S} B_{k_1}\hsp
H_I |S k_1\rangle_0=\frac{\sqrt{\lambda}}{4 \omega_S} \int \frac{d k_2}{2 \pi} \frac{V_{S,k_2,-k_1}}{\omega_{k_1}}\left|k_2\right\rangle_0 .
\end{equation}
At leading order, a finite time evolution then yields
\beq
e^{-iH\p t}|Sk_1\rangle_0\bigg|_{O(\sl)}=\frac{-i\sqrt{\lambda}}{4\omega_{S}\ok1} \pin{k_2}V_{S,k_2,-k_1}e^{-\frac{it}{2}(\omega_{k_1}+\os+\ok{2})}\frac{{\rm sin}\left[\left(\frac{\ok{1}+\os-\ok{2}}{2}\right)t\right]}{(\ok{1}+\os-\ok{2})/2}|k_2\rangle_0. \label{eva}
\eeq

This process is only on shell if $k_2=\pm {\kia}$ where we now define ${\kia}$ differently from the case of Stokes scattering in Sec.~\ref{stokessez}
\beq
{\oka}=\ok{2}-\os\hsp {\kia}>0.
\eeq

At large times, only the on-shell $k_2$ values contribute as
\beq
\lim{t\rightarrow\infty}\frac{{\rm sin}\left[\left(\frac{\ok{1}+\os-\ok{2}}{2}\right)t\right]}{(\ok{1}+\os-\ok{2})/2}=\left(\frac{{\oka}}{{\kia}}\right)\left(2\pi\delta(k_1-{\kia})+2\pi\delta(k_1+{\kia})\right).
\eeq
Substituting this limit into Eq.~(\ref{eva}) and folding the result into our initial wave packet (\ref{inita}) we find the anti-Stokes scattered part of the state at time $t$
\bea
e^{-iH\p t}|\Phi\rangle_0\bigg|_{O(\sl)}&=&\frac{-i\sqrt{\lambda}}{4\omega_{S}} \pin{k_1}\frac{\alpha_{k_1}}{ \ok1}\pin{k_2}V_{S,k_2,-k_1}e^{-\frac{it}{2}(\omega_{k_1}+\os+\ok{2})}\frac{{\rm sin}\left[\left(\frac{\ok{1}+\os-\ok{2}}{2}\right)t\right]}{(\ok{1}+\os-\ok{2})/2}|k_2\rangle_0\nonumber\\
&=&\frac{-i\sqrt{\lambda}}{4\omega_{S}} \pin{k_2}e^{-i\ok{2}t}\left(\frac{1}{{\kia}}\right)\left(\alpha_{{\kia}}V_{S,k_2,-{\kia}}+\alpha_{-{\kia}}V_{S,k_2,{\kia}}
\right)|k_2\rangle_0\nonumber\\
&=&\frac{-i\sigma\sqrt{\pi\lambda}}{2\os } \pin{k_2}e^{ix_0(k_0-{\kia})}e^{-\sigma^2(k_0-{\kia})^2}e^{-i{\ok{2}}t}\left(\frac{\tilde{V}_{S,k_2,-{\kia}}}{{\kia}}\right)|k_2\rangle_0
\eea
which is summarized by the matrix elements
\beq
\frac{{}_0\langle k_2|e^{-iH\p t}|\Phi\rangle_0}{{}_0\langle 0\vac_0}=\frac{-i\sigma\sqrt{\pi\lambda}}{4\os\ok{2}{\kia}} e^{ix_0(k_0-{\kia})}e^{-\sigma^2(k_0-{\kia})^2}e^{-i\ok{2}t}\tilde{V}_{S,k_2,-{\kia}}.
\eeq
In the limit $\sigma\rightarrow\infty$, in which the initial meson wave packet is monochromatic, this reduces to
\bea
\left|\frac{{}_0\langle k_2|e^{-iH\p t}|\Phi\rangle_0}{{}_0\langle 0\vac_0}\right|^2&=&\frac{\sigma^2\pi\lambda}{16\os^2\ok{2}^2{\kia}^2}\left|\tilde{V}_{S,k_2,-{\kia}}\right|^2e^{-2\sigma^2(k_0-{\kia})^2}\\
&=&\frac{\sigma\pi^{3/2}\lambda}{16\sqrt{2}\os^2\ok{2}^2{\kia}^2}\left|\tilde{V}_{S,k_2,-{\kia}}\right|^2\delta({\kia}-k_0).\nonumber
\eea

We want to calculate the probability that the final state has one ground state kink and one meson.  Such states are preserved by the projector
\beq
\mathcal{P}=\int dk_2 \mathcal{P}_{\rm{diff}}(k_2)\hsp  \mathcal{P}_{\rm{diff}}(k_2)=\frac{1}{2\pi} \frac{2\ok{2}}{{}_0\langle 0\vac_0} |k_2\rangle_0{}_0\langle k_2|.
\eeq


Including the correction factor for the norm of the initial state
\bea
\frac{{}_0\langle\Phi|\Phi\rangle_0}{{}_0\langle 0\vac_0}&=&\pink{2}\alpha_{k_1}\alpha^*_{k_2}\frac{{}_0\langle Sk_2|Sk_1\rangle_0}{{}_0\langle 0\vac_0}=\pin{k}\frac{|\alpha_k|^2}{4\os\ok{}}=\frac{1}{4\os\omega_{k_0}}\pin{k}|\alpha_k|^2\\
&=&\frac{1}{4\os\omega_{k_0}}\pin{k}\int dx\int dy g_k(x)g^*_k(y)\Phi(x)\Phi^*(y)
\nonumber\\
&=&\frac{1}{4\os\omega_{k_0}}\int dx |\Phi(x)|^2=\frac{\sigma\sqrt{\pi}}{2\sqrt{2}\os\omega_{k_0}}\nonumber
\eea
where we again used $\ok{}\sim \ok{0}$ in the last step of the first line, we find that the total probability of anti-Stokes scattering is
\bea
P_{\rm{aS}}&=&\frac{{}_0\langle\Phi|e^{iH\p t}\mathcal{P}e^{-iH\p t}|\Phi\rangle_0}{{}_0\langle\Phi|\Phi\rangle_0}=\pin{k_2} \frac{2\ok{2}}{{}_0\langle 0\vac_0}\frac{\left|{}_0\langle k_2|e^{-iH\p t}|\Phi\rangle_0\right|^2}{{}_0\langle\Phi|\Phi\rangle_0/{}_0\langle 0\vac_0}\frac{1}{{}_0\langle 0\vac_0}\\
&=&\pin{k_2}2\ok{2}\frac{\frac{\sigma\pi^{3/2}\lambda}{16\sqrt{2}\os^2\ok{2}^2{\kia}^2}\left|\tilde{V}_{S,k_2,-{\kia}}\right|^2\delta({\kia}-k_0)}{\left(\frac{\sigma\sqrt{\pi}}{2\sqrt{2}\os\omega_{k_0}}\right)}
\nonumber\\
&=&\frac{\pi\lambda\ok{0}}{4\os (\ok{0}+\os)k_0^2}\pin{k_2}\left|\tilde{V}_{S,k_2,-{\kia}}\right|^2\delta({\kia}-k_0)\nonumber\\
&=&\lambda\frac{\left|\tilde{V}_{S,\sqrt{(\ok{0}+\os)^2-m^2},-k_0}\right|^2+\left|\tilde{V}_{S,-\sqrt{(\ok{0}+\os)^2-m^2},-k_0}\right|^2
}{8\os k_0\sqrt{(\ok{0}+\os)^2-m^2}}.\nonumber
\eea
Again the first term is the probability that the outgoing meson travels in the same direction as the incoming meson.

\section{Example: $\phi^4$ Double-Well Model} \label{fsez}

\subsection{Analytic Results}

Consider the $\phi^4$ double-well model, which is defined by the potential
\beq
V(\sqrt{\lambda}\phi(x))=\frac{\lambda\phi^2(x)}{4}\left(\sqrt{\lambda}\phi(x)-\sqrt{2}m\right)^2
.
\eeq
It has a single shape mode, with frequency
\beq
\os=\sqrt{3}\beta\hsp \beta=\frac{m}{2}.
\eeq
The normal modes are
\bea
\g_k(x)&=&\frac{e^{-ikx}}{\ok{} \sqrt{k^2+\b^2}}\left[k^2-2\b^2+3\b^2\sech^2(\b x)-3i\b k\tanh(\b x)\right]\label{nmode}\\
\g_S(x)&=&\sqrt\frac{3\b}{2}\tanh(\b x)\sech(\b x)\hsp
\g_B(x)=\frac{\sqrt{3\b}}{2}\sech^2(\b x)\nonumber
\eea
leading to
\beq
V_{Sk_1k_2}=\pi \frac{3\sqrt{3}}{8}\frac{\left(17\b^4-(\ok1^2-\ok2^2)^2\right)(\b^2+k_1^2+k_2^2)+8\b^2k_1^2k_2^2}{\b^{3/2}\ok1\ok2\sqrt{\b^2+k_1^2}\sqrt{\b^2+k_2^2}}\sech\left(\frac{\pi(k_1+k_2)}{2\b}\right)
\eeq
and so
\bea
\left|\tilde{V}_{S,\pm\sqrt{(\ok{0}-\os)^2-m^2},-k_0}\right|&=&\left|V_{S,\pm\sqrt{(\ok{0}-\os)^2-m^2},-k_0}\right|\\
&&\hspace{-4.8cm}=\frac{\left|\left(-10\b^2+3\os\ok0-3k_0^2\right)(k_0^2-\os\ok0+2\b^2)+\left(k_0^2-2\os\ok0+3\b^2\right)k_0^2\right|}{(\ok0-\os)\ok0\sqrt{\ok0^2-2\os\ok0}\sqrt{\b^2+k_0^2}}\nonumber\\
&&\times 3\sqrt{3\b}\pi\sech\left(\frac{\pi(\pm\sqrt{k_0^2-2\os\ok0+3\b^2}-k_0)}{2\b}\right)
\nonumber\\
&&\hspace{-4.8cm}=6\sqrt{3\b}\pi\frac{\sqrt{\ok{0}}(\ok{0}-\os)}{\sqrt{\ok0-2\os}\sqrt{\b^2+k_0^2}} \sech\left(\frac{\pi(\pm\sqrt{k_0^2-2\os\ok0+3\b^2}-k_0)}{2\b}\right).
\nonumber
\eea
The probability of Stokes scattering is then
\bea
P_{\rm{S}}&=&\frac{\ok 0(\ok 0-\os)^2}{(\ok 0-2\os)(\b^2+k_0^2)k_0\sqrt{k_0^2-2\os\ok0+3\b^2}}\nonumber\\
&&\times \frac{9\sqrt{3}\pi^2 \lambda}{2} \left[ {\rm{sech}}{}^2\left({\frac{\pi(\sqrt{k_0^2-2\os\ok0+3\b^2}-k_0)}{2\b}}\right)\nonumber\right.\\
&&+\left.{\rm{sech}}{}^2\left({\frac{\pi(\sqrt{k_0^2-2\os\ok0+3\b^2}+k_0)}{2\b}}\right)
\right].
\eea

The first sech term is the probability that the outgoing meson continues in the same direction as the initial meson, while the second is the probability that it travels in the opposite direction.  The ratio of these two possibilities is just the ratio of the two sech terms.

At this order, the energy shift in the meson sector is exactly $\os$.  However the total momentum of the mesons is not conserved.  Far from the kink, the initial meson momentum is $k_0$ and the final momentum is $\sqrt{k_0^2-2\os\ok0+3\b^2}$.  The arguments of the sech terms are the momentum transfers between the mesons and the kink in the case of forward and backward scattering.

Similarly, in the case of anti-Stokes scattering
\bea
\left|\tilde{V}_{S,\pm \sqrt{(\ok{0}+\os)^2-m^2},-k_0}\right|&=&\left|V_{S,\pm \sqrt{(\ok{0}+\os)^2-m^2},-k_0}\right|\\
&&\hspace{-4.8cm}=\frac{\left|\left(-3k_0^2-3\os\ok 0-10\b^2\right)(k_0^2+\os\ok 0 +2\b^2)+(k_0^2+2\os\ok 0 +3\b^2)k_0^2\right|}{\ok0(\ok 0+\os)\sqrt{\ok{0}^2+2\os\ok 0}\sqrt{\b^2+k_0^2}}\nonumber\\
&&\times 3\sqrt{3\b}\pi\sech\left(\frac{\pi\left( \sqrt{k_0^2+2\os\ok 0 +3\b^2}\pm k_0\right)}{2\b}\right)
\nonumber\\
&&\hspace{-4.8cm}=6\sqrt{3\b}\pi\frac{\sqrt{\ok{0}}(\ok{0}+\os)}{\sqrt{\ok0+2\os}\sqrt{\b^2+k_0^2}} \sech\left(\frac{\pi(\pm\sqrt{k_0^2+2\os\ok0+3\b^2}-k_0)}{2\b}\right)
\nonumber
\eea
leading to a probability of
\bea
P_{\rm{aS}}&=&\frac{\ok 0(\ok 0+\os)^2}{(\ok 0+2\os)(\b^2+k_0^2)k_0\sqrt{k_0^2+2\os\ok0+3\b^2}}\nonumber\\
&&\times \frac{9\sqrt{3}\pi^2 \lambda}{2} \left[ {\rm{sech}}{}^2\left({\frac{\pi(\sqrt{k_0^2+2\os\ok0+3\b^2}-k_0)}{2\b}}\right)\nonumber\right.\\
&&+\left.{\rm{sech}}{}^2\left({\frac{\pi(\sqrt{k_0^2+2\os\ok0+3\b^2}+k_0)}{2\b}}\right)
\right].
\eea

\subsection{Numerical Results}

\begin{figure}[htbp]
\centering
\includegraphics[width = 0.45\textwidth]{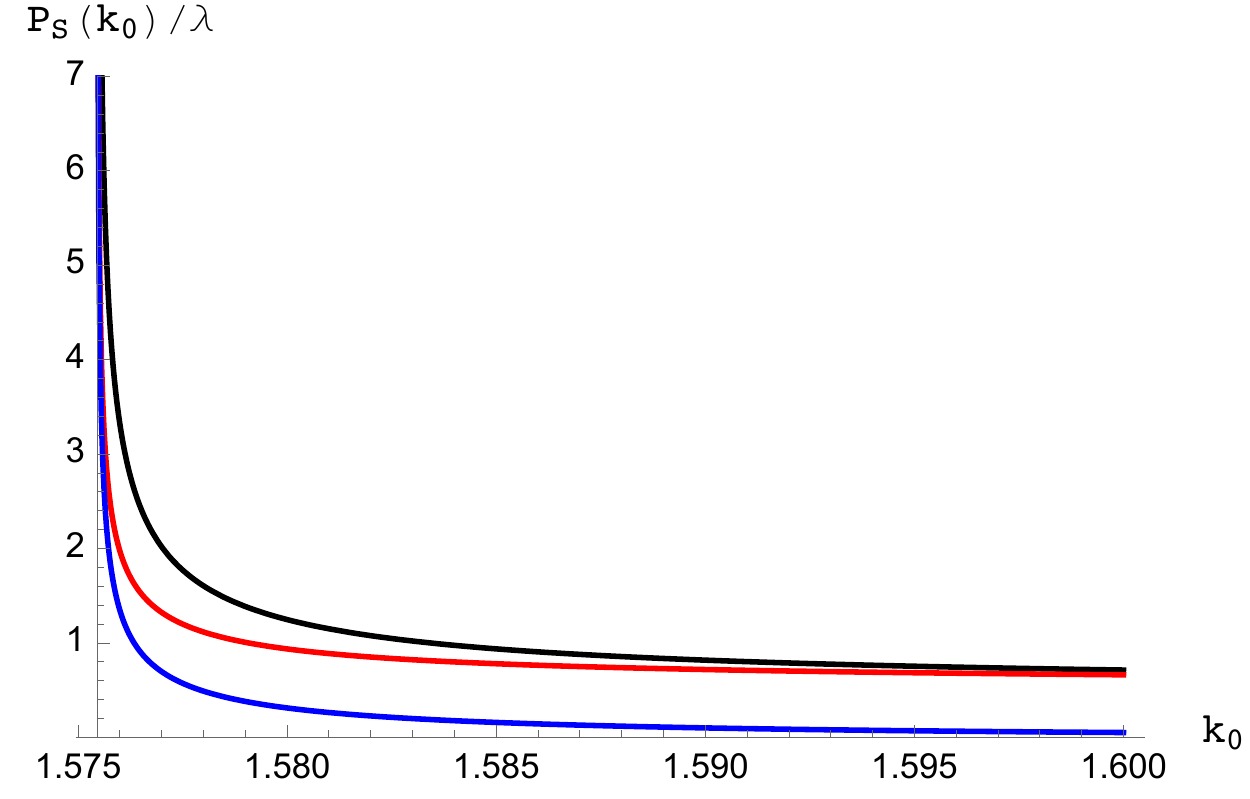}
\includegraphics[width = 0.45\textwidth]{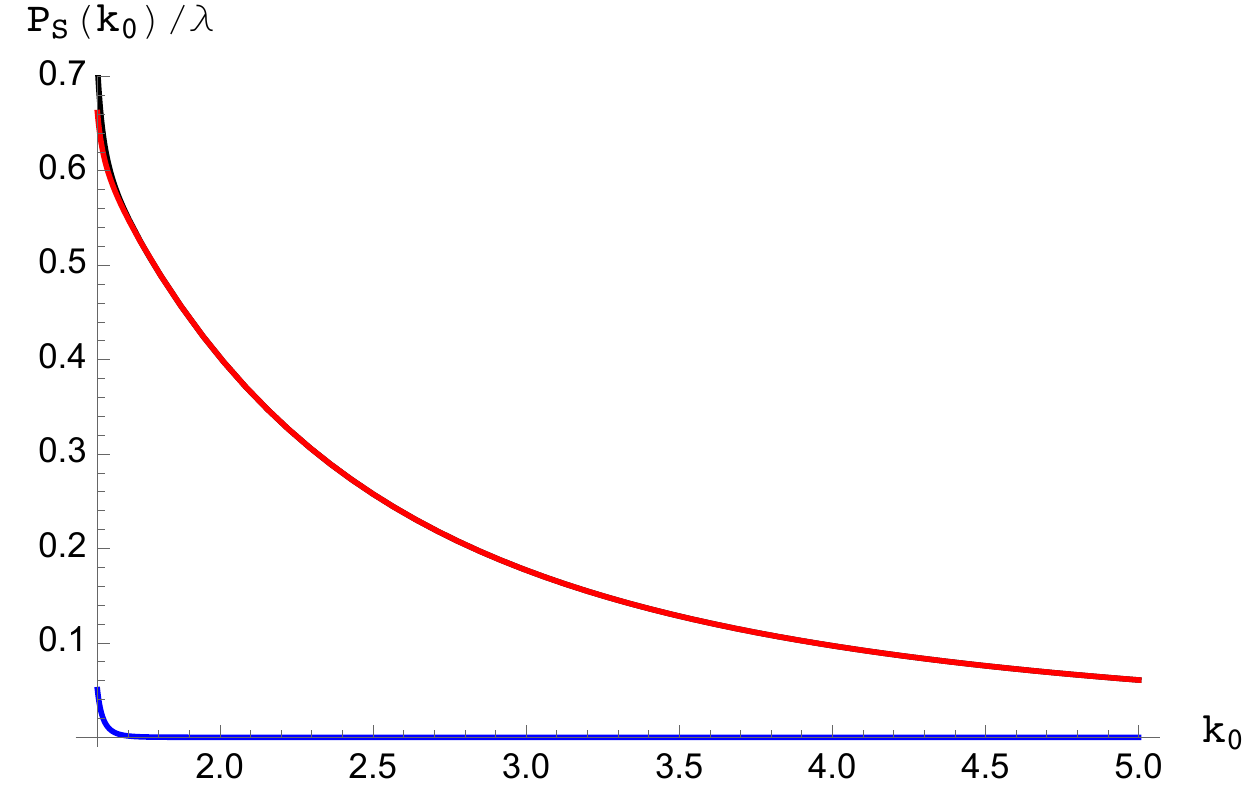}
\caption{The forward (red), backward  (blue) and total (black) probabilities $P_{S}(k_0)$ of Stokes scattering, with $m=1$. }\label{ps}
\end{figure}

\begin{figure}[htbp]
\centering
\includegraphics[width = 0.45\textwidth]{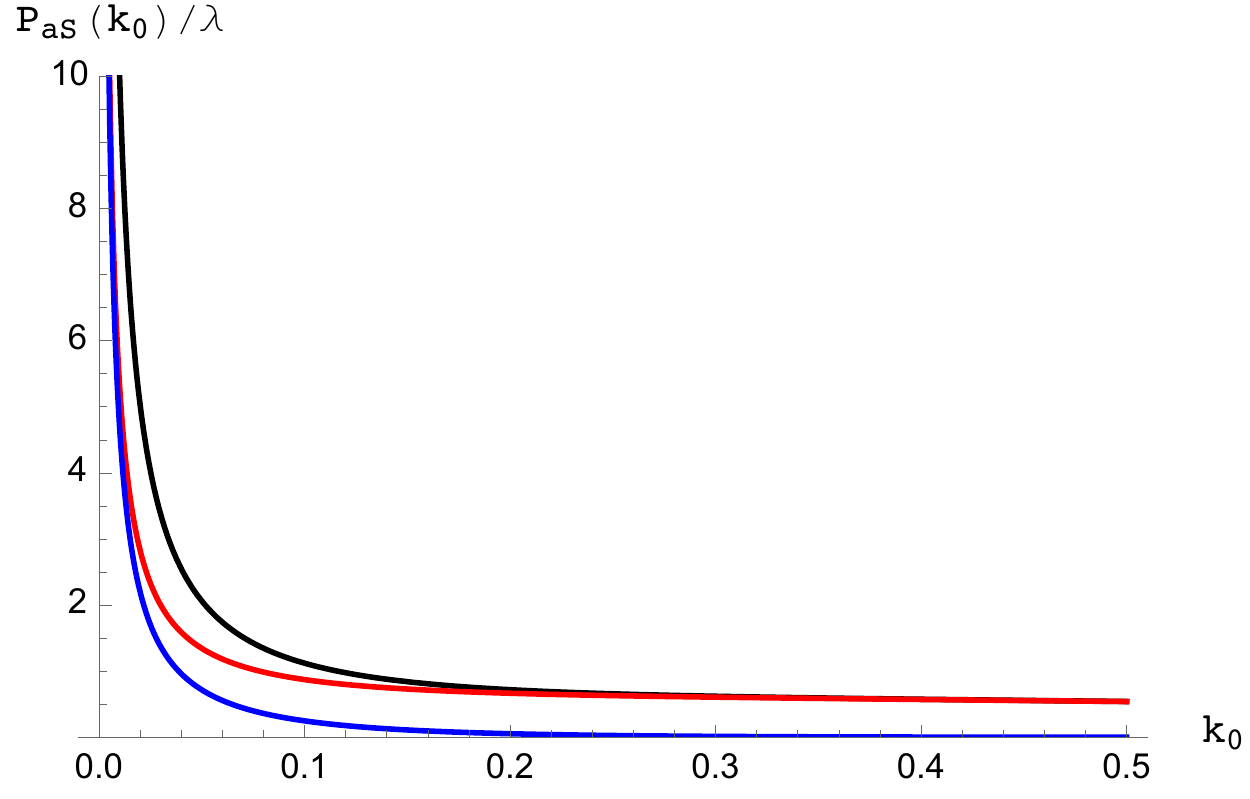}
\includegraphics[width = 0.45\textwidth]{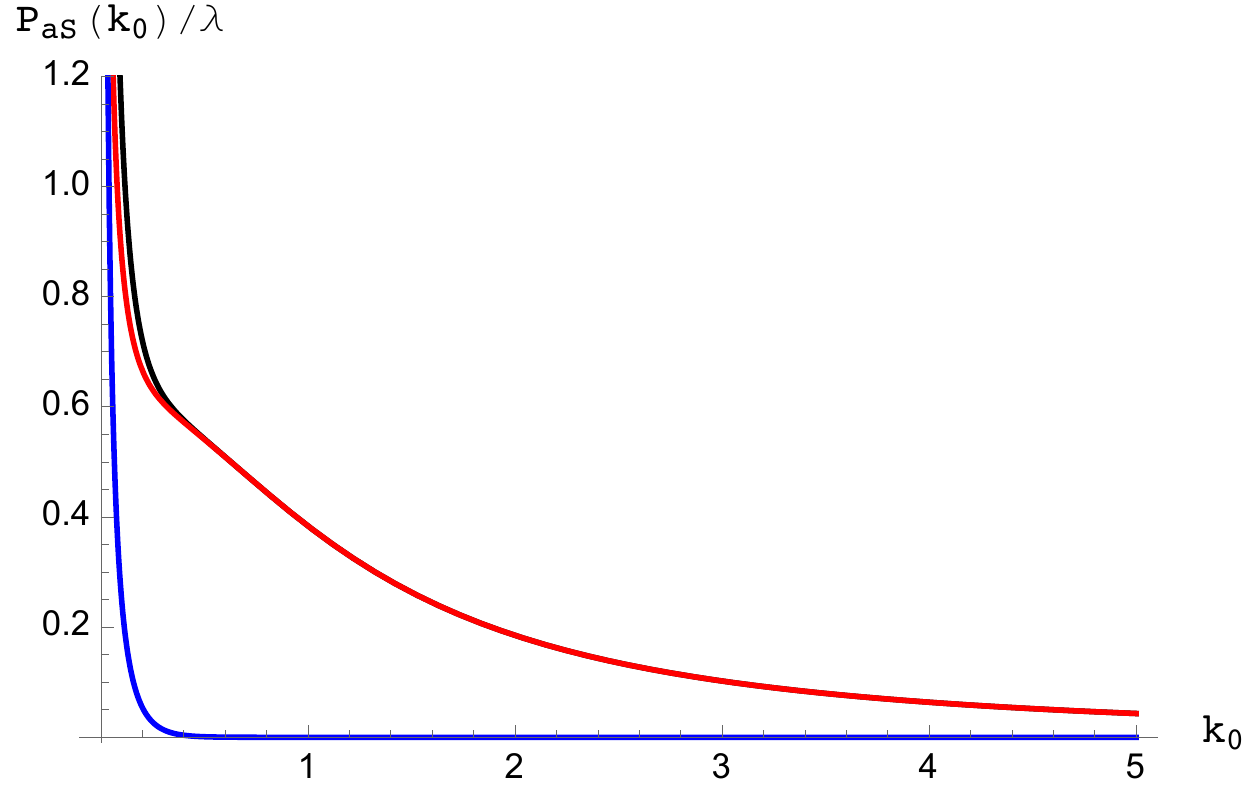}
\caption{The forward (red), backward (blue) and total (black) probabilities $P_{aS}(k_0)$ of anti-Stokes scattering, with $m=1$.}\label{pas}
\end{figure}

The probabilities depend on the dimensionless coupling $\lambda/m^2$ as well as the dimensionless momentum $k_0/m$.  We have fixed our units such that the meson mass, far from a kink, is $m=1$.  The probabilities of Stokes scattering on a ground state kink and anti-Stokes scattering on an excited kink are plotted in Figs.~\ref{ps} and \ref{pas} respectively.  

Note that Stokes scattering is only energetically allowed for a sufficiently high initial momentum, whereas the probability of anti-Stokes scattering diverges at small momenta.  Of course, once the probability, not divided by $\lambda$, is of order unity, higher order corrections dominate.  In both cases, close to the threshold, backward and forward scattering become equally probable. 

\begin{figure}[htbp]
\centering
\includegraphics[width = 0.65\textwidth]{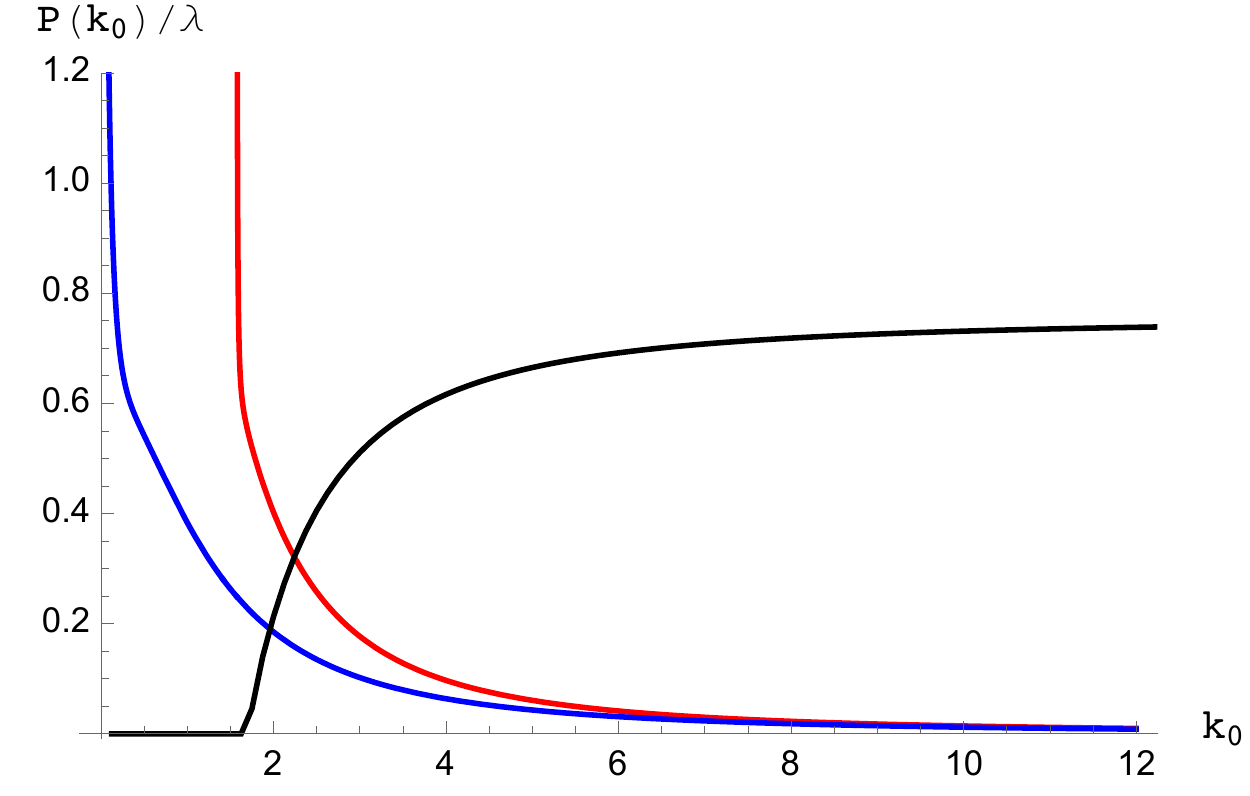}
\caption{The total probability of meson multiplication (black) from Ref.~\cite{memult}, plotted against the probability of Stokes  (red)  and anti-Stokes (blue)  scattering.}\label{pfig}
\end{figure}

In Fig.~\ref{pfig} we compare the total probabilities of these processes to that of the only other inelastic process allowed at this order, meson multiplication \cite{memult}.  This is the process in which a kink and a meson collide, yielding a kink and two mesons.  While the probabilities of Stokes and anti-Stokes scattering tend to zero for large initial momenta, the probability of meson multiplication tends to a constant.  In particular, we see that (anti)Stokes scattering dominates for low initial meson momenta, while meson multiplication dominates at higher momenta, with a cross-over when the initial momentum is about twice the meson mass.

\section{Remarks}

At order $O(\lambda)$, the inelastic scattering of a quantum kink and fundamental meson is now fully understood.  There are three allowed processes.  First, in meson multiplication, the meson may split in two.  Second, if the kink is in its ground state, then when the meson interacts it may excite a shape mode.  Finally, if a shape mode is initially excited, then when the meson interacts it may de-excite the shape mode.  The first interaction dominates at high energies, while the others become very large near their low energy thresholds.

We have always begun with an eigenstate of the free Hamiltonian, and measured the state in an eigenstate of the free Hamiltonian.  This involves matrix elements which are formally infinite, as one must integrate over all possible positions of the center of mass in the compact space.  However, the same matrix elements appear in the numerator and denominator, and so fortunately they cancel.  In a companion paper \cite{menorm} we treat such ratios more carefully, dividing by the translation symmetry so that the numerator and denominator are both finite. We find that indeed there are corrections to the results obtained via a naive cancellation.  However these corrections are suppressed by a power of $\lambda$, and so are not relevant here.  If one wishes to compute loop corrections, however, the corrections found in Ref.~\cite{menorm} must be included as they enter at the same order.


While we maintain that kink-meson scattering is of intrinsic interest, it also contributes to our understanding of the interactions of kinks with their environment.  In the linear regime, we expect this interaction to be dominated by just the processes described above.  Shortly beyond the linear regime, on the other hand, there will be processes which are of higher order in the amplitude of the radiation, such as meson fusion \cite{tomrad1,tomrad2,tomrad3}.  These become more relevant as one transitions to the classical regime.  In the near future we would like to understand such higher order processes in the quantum theory.

\section* {Acknowledgement}

\noindent
JE is supported by NSFC MianShang grants 11875296 and 11675223. HL acknowledges the support from CAS-DAAD Joint Fellowship Programme for Doctoral students of UCAS.


\begin{thebibliography}{99}

\bibitem{kklan}
Y.~Zhong, X.~L.~Du, Z.~C.~Jiang, Y.~X.~Liu and Y.~Q.~Wang,
``Collision of two kinks with inner structure,''
JHEP \textbf{02} (2020), 153
doi:10.1007/JHEP02(2020)153
[arXiv:1906.02920 [hep-th]].

\bibitem{kk0}
H.~Yan, Y.~Zhong, Y.~X.~Liu and K.~i.~Maeda,
``Kink-antikink collision in a Lorentz-violating $\phi^4$ model,''
Phys. Lett. B \textbf{807} (2020), 135542
doi:10.1016/j.physletb.2020.135542
[arXiv:2004.13329 [hep-th]].

\bibitem{kk1}
M.~Mohammadi and E.~Momeni,
``Scattering of kinks in the B\ensuremath{\varphi}4 model,''
Chaos Solitons and Fractals: the interdisciplinary journal of Nonlinear Science and Nonequilibrium and Complex Phenomena \textbf{165} (2022), 112834
doi:10.1016/j.chaos.2022.112834
[arXiv:2207.00655 [nlin.CD]].

\bibitem{kk2}
I.~Takyi, S.~Gyampoh, B.~Barnes, J.~Ackora-Prah and G.~A.~Okyere,
``Kink Collision in the Noncanonical $\varphi^{6}$ Model: A Model with Localized Inner Structures,''
[arXiv:2209.05902 [hep-th]].


\bibitem{csw}
D.~K. Campbell, J.~F. Schonfeld and C.~A. Wingate,
``Resonance structure in kink-antikink interactions in $\phi^4$ theory,"
Physica \textbf{D9} (1983) 1.

\bibitem{chaosmod}
N.~S.~Manton, K.~Oles, T.~Romanczukiewicz and A.~Wereszczynski,
``Collective Coordinate Model of Kink-Antikink Collisions in \ensuremath{\phi}4 Theory,''
Phys. Rev. Lett. \textbf{127} (2021) no.7, 071601
doi:10.1103/PhysRevLett.127.071601
[arXiv:2106.05153 [hep-th]].

\bibitem{int}
A.~Moradi Marjaneh, F.~C.~Simas and D.~Bazeia,
``Collisions of kinks in deformed \ensuremath{\varphi}4 and \ensuremath{\varphi}6 models,''
Chaos Solitons and Fractals: the interdisciplinary journal of Nonlinear Science and Nonequilibrium and Complex Phenomena \textbf{164} (2022), 112723
doi:10.1016/j.chaos.2022.112723
[arXiv:2207.00835 [hep-th]].

\bibitem{int2}
A.~Alonso-Izquierdo, D.~Migu\'elez-Caballero, L.~M.~Nieto and J.~Queiroga-Nunes,
``Wobbling kinks in a two-component scalar field theory: Interaction between shape modes,''
[arXiv:2207.10989 [hep-th]].

\bibitem{doreyf6}
P.~Dorey, K.~Mersh, T.~Romanczukiewicz and Y.~Shnir,
``Kink-antikink collisions in the $\phi^6$ model,''
Phys. Rev. Lett. \textbf{107} (2011), 091602
doi:10.1103/PhysRevLett.107.091602
[arXiv:1101.5951 [hep-th]].

\bibitem{f622}
C.~Adam, P.~Dorey, A.~Garcia Martin-Caro, M.~Huidobro, K.~Oles, T.~Romanczukiewicz, Y.~Shnir and A.~Wereszczynski,
``Multikink scattering in the $\phi^6$ model revisited,''
[arXiv:2209.08849 [hep-th]].

\bibitem{sfal21}
C.~Adam, D.~Ciurla, K.~Oles, T.~Romanczukiewicz and A.~Wereszczynski,
``Sphalerons and resonance phenomenon in kink-antikink collisions,''
Phys. Rev. D \textbf{104} (2021) no.10, 105022
doi:10.1103/PhysRevD.104.105022
[arXiv:2109.01834 [hep-th]].

\bibitem{col22}
C.~Adam, P.~Dorey, A.~Garcia Martin-Caro, M.~Huidobro, K.~Oles, T.~Romanczukiewicz, Y.~Shnir and A.~Wereszczynski,
``Multikink scattering in the $\phi^6$ model revisited,''
[arXiv:2209.08849 [hep-th]].


\bibitem{muri}
C.~Adam, K.~Oles, T.~Romanczukiewicz and A.~Wereszczynski,
``Spectral Walls in Soliton Collisions'',
Phys. Rev. Lett. \textbf{122} (2019) no.24, 241601
doi:10.1103/PhysRevLett.122.241601
[arXiv:1903.12100].

\bibitem{fmuri}
J.~G.~F.~Campos, A.~Mohammadi, J.~M.~Queiruga, A.~Wereszczynski and W.~J.~Zakrzewski,
``Fermionic spectral walls in kink collisions,''
[arXiv:2211.07754 [hep-th]].

\bibitem{chris}
J.~Evslin, C.~Halcrow, T.~Romanczukiewicz and A.~Wereszczynski,
``Spectral walls at one loop,''
Phys. Rev. D \textbf{105} (2022) no.12, 125002
doi:10.1103/PhysRevD.105.125002
[arXiv:2202.08249 [hep-th]].

\bibitem{osc}
H.~Segur and M.~D.~Kruskal,
``Nonexistence of Small Amplitude Breather Solutions in $\phi^4$ Theory,''
Phys. Rev. Lett. \textbf{58} (1987), 747-750
doi:10.1103/PhysRevLett.58.747

\bibitem{osc3d}
G.~Fodor, P.~Forgacs, P.~Grandclement and I.~Racz,
``Oscillons and Quasi-breathers in the phi**4 Klein-Gordon model,''
Phys. Rev. D \textbf{74} (2006), 124003
doi:10.1103/PhysRevD.74.124003
[arXiv:hep-th/0609023 [hep-th]].

\bibitem{quantosc}
M.~P.~Hertzberg,
``Quantum Radiation of Oscillons,''
Phys. Rev. D \textbf{82} (2010), 045022
doi:10.1103/PhysRevD.82.045022
[arXiv:1003.3459 [hep-th]].

\bibitem{hayashi1}
A.~Hayashi, S.~Saito and M.~Uehara,
``Pion - nucleon scattering in the Skyrme model and the P wave Born amplitudes,''
Phys. Rev. D \textbf{43} (1991), 1520-1531
doi:10.1103/PhysRevD.43.1520

\bibitem{hayashi2}
A.~Hayashi, S.~Saito and M.~Uehara,
``Pion - nucleon scattering in the soliton model,''
Prog. Theor. Phys. Suppl. \textbf{109} (1992), 45-72
doi:10.1143/PTPS.109.45


\bibitem{mekink}
J.~Evslin,
``Manifestly Finite Derivation of the Quantum Kink Mass,''
JHEP \textbf{11} (2019), 161
doi:10.1007/JHEP11(2019)161
[arXiv:1908.06710 [hep-th]].

\bibitem{me2loop}
J.~Evslin and H.~Guo,
``Two-Loop Scalar Kinks,''
Phys. Rev. D \textbf{103} (2021) no.12, 125011
doi:10.1103/PhysRevD.103.125011
[arXiv:2012.04912 [hep-th]].

\bibitem{gjscc}
J.~L.~Gervais, A.~Jevicki and B.~Sakita,
``Collective Coordinate Method for Quantization of Extended Systems,''
Phys. Rept. \textbf{23} (1976), 281-293
doi:10.1016/0370-1573(76)90049-1

\bibitem{gj76}
J.~L.~Gervais and A.~Jevicki,
``Point Canonical Transformations in Path Integral,''
Nucl. Phys. B \textbf{110} (1976), 93-112
doi:10.1016/0550-3213(76)90422-3

\bibitem{memult}
J.~Evslin, H.~Liu and B.~Zhang,
``Kinks Multiply Mesons,''
[arXiv:2211.01794 [hep-th]].

\bibitem{wstabile}
H.~Weigel,
``Quantum Instabilities of Solitons,''
AIP Conf. Proc. \textbf{2116} (2019) no.1, 170002
doi:10.1063/1.5114153
[arXiv:1907.10942 [hep-th]].

\bibitem{cahill76}
K.~E.~Cahill, A.~Comtet and R.~J.~Glauber,
``Mass Formulas for Static Solitons,''
Phys. Lett. B \textbf{64} (1976), 283-285
doi:10.1016/0370-2693(76)90202-1

\bibitem{meip}
J.~Evslin and H.~Liu,
``A Reduced Inner Product for Kink States,''
[arXiv:2212.10344 [hep-th]].

\bibitem{menorm}
J.~Evslin and H.~Liu,
``A Reduced Inner Product for Kink States,''
[arXiv:2212.10344 [hep-th]].

\bibitem{tomrad1}
T.~Romanczukiewicz,
``Interaction between kink and radiation in phi**4 model,''
Acta Phys. Polon. B \textbf{35} (2004), 523-540
[arXiv:hep-th/0303058 [hep-th]].

\bibitem{tomrad2}
T.~Romanczukiewicz,
``Interaction between topological defects and radiation,''
Acta Phys. Polon. B \textbf{36} (2005), 3877-3887

\bibitem{tomrad3}
P.~Forgacs, A.~Lukacs and T.~Romanczukiewicz,
``Negative radiation pressure exerted on kinks,''
Phys. Rev. D \textbf{77} (2008), 125012
doi:10.1103/PhysRevD.77.125012
[arXiv:0802.0080 [hep-th]].

\end{thebibliography}

\begin{thebibliography}{99}

\bibitem{csw} D.~K. Campbell, J.~F. Schonfeld and C.~A. Wingate,
``Resonance structure in kink-antikink interactions in $\phi^4$ theory,"
Physica \textbf{D9} (1983) 1.

\bibitem{osc}
H.~Segur and M.~D.~Kruskal,
``Nonexistence of Small Amplitude Breather Solutions in $\phi^4$ Theory,''
Phys. Rev. Lett. \textbf{58} (1987), 747-750
doi:10.1103/PhysRevLett.58.747

\bibitem{osc3d}
G.~Fodor, P.~Forgacs, P.~Grandclement and I.~Racz,
``Oscillons and Quasi-breathers in the phi**4 Klein-Gordon model,''
Phys. Rev. D \textbf{74} (2006), 124003
doi:10.1103/PhysRevD.74.124003
[arXiv:hep-th/0609023 [hep-th]].

\bibitem{doreyf6}
P.~Dorey, K.~Mersh, T.~Romanczukiewicz and Y.~Shnir,
``Kink-antikink collisions in the $\phi^6$ model,''
Phys. Rev. Lett. \textbf{107} (2011), 091602
doi:10.1103/PhysRevLett.107.091602
[arXiv:1101.5951 [hep-th]].

\bibitem{multex22a}
F.~C.~Simas, K.~Z.~Nobrega, D.~Bazeia and A.~R.~Gomes,
``Degeneracy and kink scattering in a two coupled scalar field model in $(1,1)$ dimensions,''
[arXiv:2201.03372 [hep-th]].

\bibitem{multex22b}
A.~Moradi Marjaneh, F.~C.~Simas and D.~Bazeia,
``Collisions of kinks in deformed  \ensuremath{\varphi}${}^4$ and \ensuremath{\varphi}${}^6$ models,''
Chaos Solitons and Fractals: the interdisciplinary journal of Nonlinear Science and Nonequilibrium and Complex Phenomena \textbf{164} (2022), 112723
doi:10.1016/j.chaos.2022.112723
[arXiv:2207.00835 [hep-th]].

\bibitem{sfal21}
C.~Adam, D.~Ciurla, K.~Oles, T.~Romanczukiewicz and A.~Wereszczynski,
``Sphalerons and resonance phenomenon in kink-antikink collisions,''
Phys. Rev. D \textbf{104} (2021) no.10, 105022
doi:10.1103/PhysRevD.104.105022
[arXiv:2109.01834 [hep-th]].

\bibitem{col22}
C.~Adam, P.~Dorey, A.~Garcia Martin-Caro, M.~Huidobro, K.~Oles, T.~Romanczukiewicz, Y.~Shnir and A.~Wereszczynski,
``Multikink scattering in the $\phi^6$ model revisited,''
[arXiv:2209.08849 [hep-th]].

\bibitem{mm}
N.~S.~Manton and H.~Merabet,
``$\phi^4$ kinks: Gradient flow and dynamics,''
Nonlinearity \textbf{10} (1997), 3
doi:10.1088/0951-7715/10/1/002
[arXiv:hep-th/9605038 [hep-th]].

\bibitem{tomrad1}
T.~Romanczukiewicz,
``Interaction between kink and radiation in phi**4 model,''
Acta Phys. Polon. B \textbf{35} (2004), 523-540
[arXiv:hep-th/0303058 [hep-th]].

\bibitem{tomrad2}
T.~Romanczukiewicz,
``Interaction between topological defects and radiation,''
Acta Phys. Polon. B \textbf{36} (2005), 3877-3887

\bibitem{tomrad3}
P.~Forgacs, A.~Lukacs and T.~Romanczukiewicz,
``Negative radiation pressure exerted on kinks,''
Phys. Rev. D \textbf{77} (2008), 125012
doi:10.1103/PhysRevD.77.125012
[arXiv:0802.0080 [hep-th]].

\bibitem{mekink}
J.~Evslin,
``Manifestly Finite Derivation of the Quantum Kink Mass,''
JHEP \textbf{11} (2019), 161
doi:10.1007/JHEP11(2019)161
[arXiv:1908.06710 [hep-th]].

\bibitem{me2loop}
J.~Evslin and H.~Guo,
``Two-Loop Scalar Kinks,''
Phys. Rev. D \textbf{103} (2021) no.12, 125011
doi:10.1103/PhysRevD.103.125011
[arXiv:2012.04912 [hep-th]].

\bibitem{gjscc}
J.~L.~Gervais, A.~Jevicki and B.~Sakita,
``Collective Coordinate Method for Quantization of Extended Systems,''
Phys. Rept. \textbf{23} (1976), 281-293
doi:10.1016/0370-1573(76)90049-1

\bibitem{gj76}
J.~L.~Gervais and A.~Jevicki,
``Point Canonical Transformations in Path Integral,''
Nucl. Phys. B \textbf{110} (1976), 93-112
doi:10.1016/0550-3213(76)90422-3

\bibitem{wstabile}
H.~Weigel,
``Quantum Instabilities of Solitons,''
AIP Conf. Proc. \textbf{2116} (2019) no.1, 170002
doi:10.1063/1.5114153
[arXiv:1907.10942 [hep-th]].

\bibitem{cahill76}
K.~E.~Cahill, A.~Comtet and R.~J.~Glauber,
``Mass Formulas for Static Solitons,''
Phys. Lett. B \textbf{64} (1976), 283-285
doi:10.1016/0370-2693(76)90202-1

\bibitem{meip}
J.~Evslin and H.~Liu,
``A Reduced Inner Product for Kink States,''
[arXiv:2212.10344 [hep-th]].

\bibitem{memult}
J.~Evslin, H.~Liu and B.~Zhang,
``Kinks Multiply Mesons,''
[arXiv:2211.01794 [hep-th]].

\bibitem{phi42loop}
J.~Evslin,
``\ensuremath{\phi^4} kink mass at two loops,''
Phys. Rev. D \textbf{104} (2021) no.8, 085013
doi:10.1103/PhysRevD.104.085013
[arXiv:2104.07991 [hep-th]].

\bibitem{merif}
J.~Evslin and H.~Liu,
``Quantum Reflective Kinks,''
[arXiv:2210.12725 [hep-th]].

\end{thebibliography}
\end{document}